\documentclass[aps,twocolumn,epsfig,amsfonts,amsmath,amssymb,prb]{revtex4}
\usepackage{graphicx}
\usepackage{graphics}
\begin{document}
\title{Intermediate range chemical ordering of cations in simple molten alkali halides}

\author{M Salanne$^{1,2,*}$, C Simon$^{1,2}$, P Turq$^{1,2}$ and P A Madden$^3$}
\address{$^1$UPMC Univ Paris 06, UMR 7612, LI2C, F-75005, Paris, France}
\address{$^2$CNRS, UMR 7612, LI2C, F-75005, Paris, France}
\address{$^3$School of  Chemistry, University of Edinburgh, Edinburgh EH9 3JJ, UK}
\address{$^*$mathieu.salanne@upmc.fr}

\begin{abstract}
The presence of first sharp diffraction peaks in the partial  structure factors is
investigated in computer simulations of molten mixtures of alkali halides. An
intermediate range ordering appears for the Li$^+$ ions only, which is associated
with clustering of this species and is not reflected in the arrangement of other
ions. This ordering is surprising in view of the simplicity of the interionic
interations in alkali halides. The clustering reflects an incomplete mixing of the
various species on a local length scale, which can be demonstrated by studying the
{\it complementary} sub-space of cations in the corresponding pure alkali halides
by means of a void analysis.
\end{abstract}

\maketitle

Prepeaks or first sharp diffraction peaks in the diffraction structure factors of
amorphous solids or liquids are associated with intermediate-range order \cite{salmon2006a}. In simple
liquids, like hard-spheres or simple MX molten salts the structure factors $S(k)$
are dominated by a principal peak at $k_{pp}\sim \frac {2\pi}{\sigma}$, where
$\sigma$ is the nearest neighbour separation, and have only a small amplitude for
smaller values of $k$, monotonically decaying as $k\to 0$ \cite{revere1986a}. This effect, which is
captured in simple integral theories of such fluids, means that, although the
radial distribution functions of these systems do exhibit oscillatory structure out
to much larger separations than $\sigma$, it is simply a consequence of the packing
of the particles reflected in the nearest-neighbour separation. Alternatively,
viewed in reciprocal space, the density fluctuations reflected in $S(k)$  are
suppressed for $k< k_{pp}$ by packing effects in van der Waals atomic liquids and
by the combination of packing and coulomb ordering in simple molten salts.

Exceptions are found when a competing interaction affects the order.
In the atomic system phosphorus \cite{katayama2000a}, where the
competition between tendency to form local quasi-molecular tetrahedra, which
determines the nearest-neighbour separation, and the packing of these tetrahedra in
a disordered network introduces new structure on the intermediate lengthscale and
results in a substantial prepeak. Prepeaks have been most extensively studied in
the partial structure factors of network-forming MX$_2$ binary systems, like
SiO$_2$ \cite{susman1991a} and ZnCl$_2$ \cite{salmon2005a} where, again, they reflect a
competition between the formation of local coordination polyhedra (in these cases,
tetrahedral SiO$_4$ and ZnCl$_4$ units) and the arrangement of these units in three
dimensions which is influenced by the preferred value for the cation-anion-cation
bond angle \cite{wilson1994a,wilson1998a}. Prominent prepeaks are also seen in MX$_3$
systems, for similar reasons \cite{hutchinson1999a}. In these cases it has been found
useful to associate the position of the prepeak with the arrangement of voids in
the spatial distribution of the species of interest \cite{elliott1991a,wilson1998a}; the
ordering of the voids reflects the inhomogeneities introduced in the arrangements of atoms
due to these competing structural influences. The void-void structure factor shows
a principal peak which coincides with the position of the prepeak in the atomic
structure factor. In the simple systems which lack prepeaks, the void and atomic
principal peaks coincide.

In this article we are concerned with the extent to which such ideas can be
transferred to discuss the structure of ternary mixtures (and also of binary
mixtures of metals which might also be thought of as ternary, with the valence
electrons thought of as the third structural component). Here ``prepeaks" may be
observed for essentially trivial reasons, and such cases should be discounted from
consideration. If, for example, the network-former ZnCl$_2$ is mixed with a
network-breaker, like RbCl, to form a ternary system, the network is broken up into
ZnCl$_4^{2-}$ molecular ion units if more than two moles of RbCl are combined with
one of ZnCl$_2$. A first sharp diffraction peak is now seen in the Zn-Zn partial
structure factor \cite{allen1992a} but it arises because the liquid should be considered
as a simple unassociated mixture of Rb$^+$ ions and ZnCl$_4^{2-}$ ions and the
``prepeak" can be associated with the principal peak of the partial structure factor of
the latter \cite{wilson1994a}. This structure can be understood from simple molten
salt theories. On the other hand, in apparently similar cases, such as when the
structure-breaker Na$_2$O is added to SiO$_2$ at equal mole fractions (thereby
allowing only incomplete network breakdown) the Na-Na partial structure factor
shows a prepeak and the associated lengthscale corresponds to the distance between
Na-containing channels in the remnants of the SiO$_2$ network \cite{meyer2004a}.
Since these channels are associated with Na conduction in the glassy state, in this
case the structure evident in the prepeak is clearly associated with very
significant effects on the ion dynamics. More general considerations link
non-trivial relationships between the single particle and collective dynamics to
the existence of a prepeak in the structure factor \cite{voigtmann2006a}, and these
considerations motivate, in part, our interest in the alkali halide systems.

In the cases mentioned above, the appearance of (non-trivial) prepeaks is
associated with strong interactions involving at least some components. However,
there is one system in which a prepeak has been reported where one would have
thought all the interactions were quite simple. Ribeiro \cite{ribeiro2003a}
reported a prepeak in the partial Li-Li structure factor in the alkali halide
mixtures LiF:KF and LiCl:KCl from computer simulations. Here both components are MX
molten salts, whose structures are well understood on the basis of packing and
coulomb ordering, as recaptured in simple HNC theories for example \cite{revere1986a}. We
will examine these and several related alkali halide mixtures in the present work.
Our objective is to understand how the intermediate-range order arises in
this system and to relate it to that seen in other, more familiar examples.


In his simulations, Ribeiro used rigid ion interaction models for all the ions.  In
recent years we have shown that taking polarization effects into account is
important to obtain a description of the physical properties of molten
salts\cite{salanne2006a} which is \emph{transferable} between the pure salts and
their mixtures. Even for the alkali halides, where \emph{effective} pair potentials
reproduce many properties remarkably well, polarizable potentials are necessary to
reproduce {\it ab initio} derived forces accurately \cite{madden2006a} and therefore to
faithfully represent the real interactions. We therefore undertook new simulations
of mixtures of alkali halides using polarizable ion models which have been derived
by force-fitting from first-principles \cite{madden2006a}, extensively tested, and
proven to give a good representation of the materials.

\begin{figure}
\begin{center}
\includegraphics[scale=0.3]{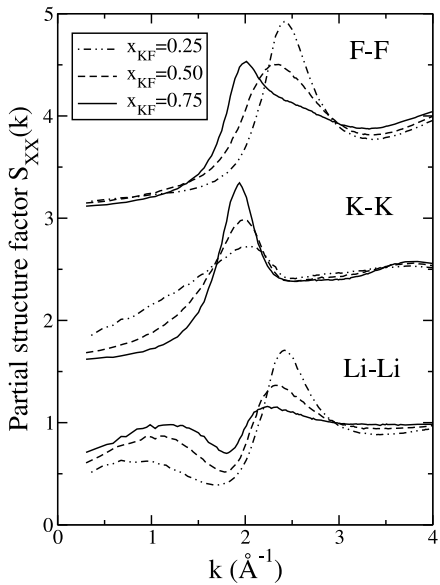}
\end{center}
\caption{\label{fig:structurefactors} F$^-$-F$^-$, K$^+$-K$^+$  and Li$^+$-Li$^+$
partial structure factors for various compositions of LiF:KF mixtures. $S_{\rm FF}
(k)$ and $S_{\rm KK} (k)$ were respectively shifted by values of 3 and 1.5.}
\end{figure}

We have performed MD simulations of KF, NaF and KCl pure melts  and of
LiF:KF ($x_{\rm KF}=$0.25, 0.50, 0.75) LiF:NaF ($x_{\rm NaF}=$0.75) and LiCl:KCl
($x_{\rm KCl}=$0.41) mixtures in the canonical ensemble. The simulation cells
contained a total of 432 ions, or 3456 ions when necessary, and were equilibrated
in the NPT ensemble at a zero pressure. The corresponding densities were in close
agreement with the experimental ones.

The like-like partial structure factors are given on figure
\ref{fig:structurefactors} for the several compositions of LiF:KF mixtures. For
K$^+$ and F$^-$ ions, only one very well defined (principal) peak is observed and
it occurs close to $k_{pp}\sim \frac {2\pi}{\sigma}$ where $\sigma$ is the
nearest-neigbour separation identified from the corresponding partial radial
distribution function.  On the other hand, $S_{\rm Li-Li}(k)$ displays two peaks
for all the compositions. The prepeak (or FSDP) is a broad peak and its intensity
rises significantly with the KF concentration of the mixture. Even if the broadness
does not allow us to determine an exact position for that peak, it clearly shifts
significantly with the change in composition, passing from
$k_{FSDP}\approx0.82$~\AA$^{-1}$ for $x_{\rm KF}=0.25$ to
$k_{FSDP}\approx1.18$~\AA$^{-1}$ for $x_{\rm KF}=0.75$. Several temperatures
ranging from 773~K to 1300~K were investigated. The corresponding number
density  of ionic pairs shifts from $2.94\times10^{-2}$~\AA$^{-3}$ ($x_{\rm KF}=0.50$ and T=773~K)
to $2.04\times10^{-2}$~\AA$^{-3}$ ($x_{\rm KF}=0.75$ and T=1300~K). For a given
composition, the positions and widths of the FSDPs remained unchanged from one
temperature to another. The peak at higher $k$ corresponds to the principal peak
and occurs at $k_{pp}\sim \frac {2\pi}{\sigma}$. The lithium ions spatial
distribution is thus characterized by the presence of a second length scale which
does not seem to affect the other ions. These findings closely parallel Ribeiro's
\cite{ribeiro2003a}.

\begin{figure}
\begin{center}
\includegraphics[scale=0.3]{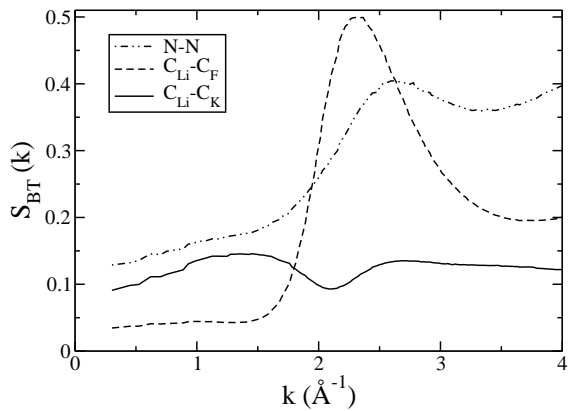}
\caption{\label{fig:bt} Bhatia-Thornton structure factors for the LiF:KF mixture at $x_{\rm KF}=0.50$ composition.}
\end{center}
\end{figure}

This behavior indicates the existence of chemical  ordering. In a binary
system, topological versus chemical ordering can be investigated by using the
Bhatia-Thornton formalism \cite{bhatia1970a,salmon2005b}. Three partial structure
factors are constructed which correspond to the number-number ($S_{NN}(k)$),
number-concentration ($S_{NC}(k)$) and concentration-concentration ($S_{CC}(k)$)
fluctuations. In a ternary mixture, the same definition can be used for
$S_{NN}(k)$:
\begin{equation}
S_{NN}(k)=\sum_{\alpha,\beta}c_\alpha c_\beta S_{\alpha\beta}(k),
\end{equation}
\noindent where $c_\alpha$ is the ionic fraction of  chemical species $\alpha$. As
regards $S_{CC}(k)$, a general expression for a multi-component fluid was developed by Bhatia\cite{bhatia1977a}, but from our perspective it is more chemically intuitive to study fluctuations for a given
pair of atoms in the same way as in a two component fluid. We have therefore determined the following partial $S_{C_\alpha C_\beta}(k)$ structure factors:
\begin{equation}
S_{C_\alpha C_\beta}(k)=c_\alpha c_\beta (S_{\alpha\alpha}(k)+S_{\beta\beta}(k)-2S_{\alpha\beta}(k)).
\end{equation}
The two concentration-concentration functions  involving Li$^+$ ions and the
number-number one have been plotted on figure \ref{fig:bt}. They give us a first
insight into the origin of the intermediate-range ordering observed for that
species. The only function which presents a peak in the low-$k$ region is
$S_{C_{\rm Li}C_{\rm K}}(k)$. This means that the Li$^+$ ions are chemically
ordered with respect to the K$^+$ but not to the F$^-$ ions.

\begin{figure}
\begin{center}
\includegraphics[scale=0.5]{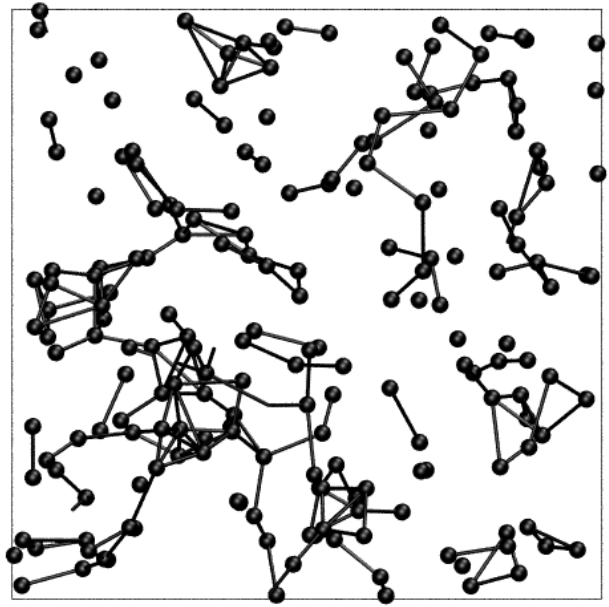}
\caption{\label{fig:Liclusters} Snapshot of the Li$^+$  ion positions in a LiF:KF
mixture ($x_{\rm KF}=0.75$). The whole cell is projected onto a plane. Pairs of Li$^+$ ions separated by a distance smaller
than the Li$^+$-Li$^+$ RDF first minima were connected by ``bonds". Note that because of the boundary conditions, some of these bonds are not appearing here.} 
\end{center}
\end{figure}

Unlike the tetrahedral network forming ionic liquids like SiO$_2$ or ZnCl$_2$, no
well-defined first coordination shell structure is formed around the lithium
cations and their fluoride ion coordination number can be either 3, 4 or 5.
Therefore, the examination of the influence of a particular ion solvation shell on
the structures occurring at a medium range scale does not give any clue to the origin of the 
intermediate range ordering. On the contrary, visual examination of the Li$^+$ ion
positions in individual configurations of the liquid reveals an heterogeneous
distribution as shown on figure \ref{fig:Liclusters}. Regions of high and low
concentrations can be identified, showing that the second length scale observed in
$S_{\rm Li-Li}(k)$ is associated with the formation of clusters of Li$^+$ ions.

Such a clustering means that the local concentration in  Li$^+$ ions fluctuates in
the liquid, revealing an incomplete mixing of LiF and KF at the nanoscopic scale.
As a consequence, the K$^+$ local concentration might be expected to fluctuate too,
but the absence of low-$k$ features in the like-like partial structure factor of
that ion shows that its distribution is homogeneous. What emerges is a picture of a
homogeneous but disordered ({\it i.e.} non network-like) matrix consisting of K$^+$
and F$^-$ ions in which clusters of Li$^+$ ions are formed. Note that the position
of the principal peak in $S_{\rm KK}$ is not shifting.  The behaviour seen
resembles closely that found by Voigtmann {\it et al} in recent experimental
studies of the binary metallic liquid Ni:Zr \cite{voigtmann2008a}. In that case,
the Ni-Ni partial structure showed a double peak structure of very similar
character to that of the Li ions in our calculation, whereas the larger host Zr
ions behaved analogously to the K$^+$ ions in our calculations: they showed a
normal simple metallic structure factor which was qualitatively reproduced by that
of hard-spheres.

In simulations of pure KF, the K$^+$ ions can be arbitrarily assigned to two
groups, which are labelled as ``hosts" and ``guests", the relative proportions of
the two groups can be varied in order to correspond to the compositions of the
LiF:KF mixtures. It is then of interest to study the {\it possible} positions at
which another species could be incorporated into the melt if the K$^+$ guests were
removed. To do this, it is possible to examine the spatial arrangement of the
complementary space of the hosts ions, {\it i.e.} of the regions where those ions
are absent.  Such ``voids" are not physical voids as they include F$^-$ anions and K$^+$ guests. They can be identified by performing a Voronoi analysis of
the host cation positions in the instantaneous liquid configuration
\cite{medvedev2005a}. The vertices of the Voronoi polyhedra in a disordered
structure define the center of a group of four atoms (here four host cations), which are mutually nearest
neighbors, and are known as the Delaunay simplices (DS) of the structure. Around
each DS a circumsphere which passes through the four atoms may be drawn and, since
no other atom center lies within the sphere, the radius $r_{c}$ of the circumsphere
gives a measure of the empty (void) space in between the atoms. From the positions
of the center of each sphere, a void-void structure factor can be calculated:
\begin{equation}
S_{\rm VV}(k)=\left\langle \frac{1}{N_{\rm V}}  \sum_{i,j=1}^{N_{\rm V}} \exp(i{\bf
k}\cdot {\bf R}^{ij}) \right\rangle.
\end{equation}
Note that $S_{\rm VV}(k)$ {\em does not} correspond to the partial structure factor
of the ``guest" K$^+$ ions in the pure melt simulation. Since these K$^+$ guests
were assigned arbitrarily ({\it i.e.} randomly), their structure factor is simply
proportional to $S_{\rm KK}(k)$ for the pure melt (plus an incoherent background).

\begin{figure}
\begin{center}
\includegraphics[scale=0.3]{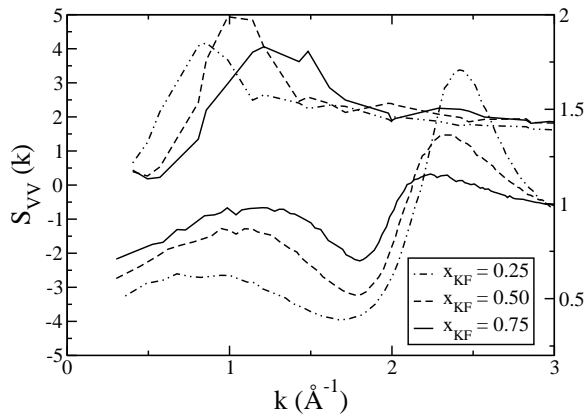}
\caption{\label{fig:sqvv} Void-void (top curves) and Li$^+$-Li$^+$ (bottom curves) partial structure factors in
LiF:KF. The void-void structure factors were determined in pure KF, where the K$^+$
ions have been separated into ``hosts" and ``guests". The voids positions were determined from the positions of the "host" ions only (see the text). The left hand ordinate axis corresponds to the void-void structure factors and the right hand ordinate axis to the Li$^+$-Li$^+$ ones.}
\end{center}
\end{figure}

In order to study the complementary space of the host K$^+$ ions in pure KF, the
positions of the voids between them have been extracted from our trajectories, and
the corresponding structure factor was computed. The results obtained for several
proportions of hosts K$^+$ ions are shown on figure \ref{fig:sqvv}, where the
Li$^+$-Li$^+$ partial structure factors of the corresponding LiF:KF compositions
are also given. One can see that the positions of the main peak of $S_{\rm VV}(k)$
and of the FSDP of $S_{\rm Li-Li}(k)$ coincide for all the compositions. This shows
that the intermediate range ordering of the Li$^+$ ions in LiF:KF occurs at the
characteristic length scale of the complementary sub-space of the corresponding
number of K$^+$ ions in pure KF, which is consistent with the view of LiF:KF
mixtures as a partial homogeneous KF matrix of which voids are filled with Li$^+$
ions.
\begin{figure}
\begin{center}
\includegraphics[scale=0.3]{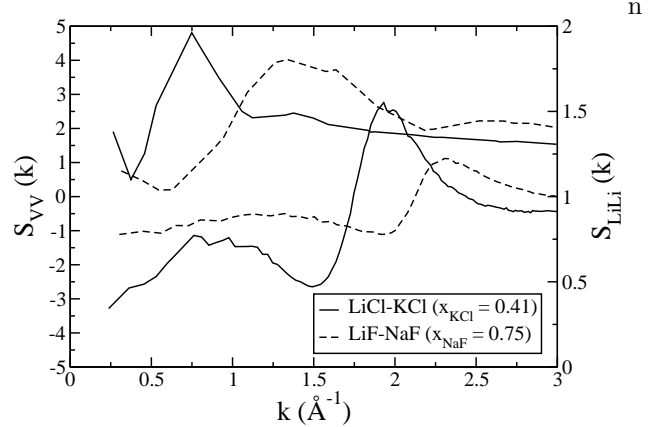}
\caption{\label{fig:sqvv2} Void-void (top curves) and  Li$^+$-Li$^+$ (bottom curves) partial structure factors.
The void-void structure factors were determined in pure KCl and NaF, where the
K$^+$ or Na$^+$ ions have been separated into ``hosts" and ``guests". The voids positions were determined from the positions of the "host" ions only (see the text). The $S_{\rm Li-Li} (k)$ were
computed in LiCl:KCl and LiF:NaF mixtures. The left hand ordinate axis corresponds to the void-void structure factors and the right hand ordinate axis to the Li$^+$-Li$^+$ ones.}
\end{center}
\end{figure}

It is natural to ask if this intermediate  range structuring is also present in
other simple mixtures of alkali halides like LiF:NaF and LiCl:KCl. The partial
structure factors computed for those systems showed exactly the same features as in
LiF:KF mixtures, {\it i.e.} the presence of a single peak in the anion structure factors and in the structure factors involving 
both K$^+$ and Na$^+$ ions, whereas $S_{\rm Li-Li}(k)$, which is shown on figure
\ref{fig:sqvv2}, presents two peaks. In LiCl:KCl, the FSDP is at $k_{FSDP}\approx
0.98$~\AA$^{-1}$\ and in LiF:NaF it appears at $ k_{FSDP}\approx 1.31$~\AA$^{-1}$. Once again,
the positions of those FSDP coincide exactly with the principal peak of the void
void structure factor of the corresponding pure host melt (KCl or NaF). This shows
that the same origin for intermediate range ordering of Li$^+$ ions is
observed in all the mixtures and that its associated length scale is dependent on
the other component of the mixture. The comparison of LiCl:KCl and LiF:KF mixture
also gives important information: because of the difference of anion, the
short-range characteristic distances are very different between those melts, and so
is the Li$^+$-Li$^+$ first neighbor distance (given by the principal peak of
$S_{\rm Li-Li}(k)$). On the contrary the intermediate range ordering of Li$^+$ ions
appears at similar length scales, which means that the knowledge of the first shell
structure does not help in understanding the clustering properties of the melt.

The intermediate range chemical ordering of  Li$^+$ ions then is a general property
of simple mixtures of lithium halides with the other alkali halides. It is due to
poor mixing properties on the nanoscopic scale, and the liquid formed consists in a
matrix of "host" ions filled with clusters of Li$^+$. The length scale associated
with those clusters is set by the void-void correlation in the pure host liquid,
and no indication is given by the real-space short-range structure. This result
shows that the formation of a network may not be a prerequisite to the formation of
those clusters, and hence of intermediate range chemical ordering in condensed
matter systems.


\end{document}